\documentclass[twocolumn,showpacs,preprintnumbers,amsmath,amssymb]{revtex4}

\usepackage{graphicx}
\usepackage{dcolumn}
\usepackage{bm}
\usepackage{epsfig}
\usepackage{amssymb}
\usepackage{amsmath}
\def\Mpl{M_{\mathrm{Pl}}}
\def\fun#1#2{\lower3.6pt\vbox{\baselineskip0pt\lineskip.9pt
  \ialign{$\mathsurround=0pt#1\hfil##\hfil$\crcr#2\crcr\sim\crcr}}}
\def\VEV#1{\left\langle #1\right\rangle}

\begin{document}

\title{Direct detection of the inflationary gravitational wave background}

\author{Tristan L. Smith$^1$, Marc Kamionkowski$^1$, and Asantha
Cooray$^{1,2}$}
\affiliation{$^1$California Institute of Technology, Mail Code
130-33, Pasadena, CA 91125}
\affiliation{$^2$Department of Physics and Astronomy, University
of California, Irvine, Irvine, CA 92697}

\date{\today}
   
\begin{abstract}
Inflation generically predicts a stochastic background of
gravitational waves over a broad range of frequencies, from
those accessible with cosmic microwave background (CMB)
measurements, to those accessible directly with
gravitational-wave detectors, like NASA's Big-Bang Observer
(BBO) or Japan's Deci-Hertz Interferometer Gravitational-wave
Observer (DECIGO), both currently under study.  Here we investigate
the detectability of the inflationary gravitational-wave
background at BBO/DECIGO frequencies.  To do so, we survey a
range of slow-roll inflationary models consistent
with constraints from the CMB and large-scale structure (LSS).  We go
beyond the usual assumption of power-law power spectra,
which may break down given the 16 orders of magnitude in frequency
between the CMB and direct detection, and solve instead the
inflationary dynamics for four classes of inflaton potentials.
Direct detection is possible in a variety of
inflationary models, although probably not in any in which the
gravitational-wave signal does not appear in the CMB polarization.
However, direct detection by BBO/DECIGO can help
discriminate between inflationary models that have the
same slow-roll parameters at CMB/LSS scales.
\end{abstract}


\maketitle

\section{Introduction}

Long a subject of theoretical speculation, inflation
\cite{Guth81,Albrecht:1982wi,Linde:1981mu} has now,
with the advent of precise cosmic microwave background (CMB)
measurements
\cite{Kamionkowski:1999qc, deBernardis:2000gy, Miller:1999qz,
Hanany:2000qf, Halverson:2001yy, Mason:2002tm, Benoit:2002mm,
Goldstein:2002gf, Spergel:2003cb},
become an empirical science.  The
concordance of the measurements with the inflationary
predictions of a flat Universe and a nearly scale-invariant
spectrum of primordial density perturbations
\cite{Guth82,Bardeen:1983qw,Hawking:1982cz,Linde:1982uu} is at least
suggestive and warrants further tests of inflation.  Among the
predictions of inflation yet to be tested is a stochastic
gravitational-wave background with a nearly scale-invariant
spectrum \cite{Starobinsky:1979ty,Abbott84,Starobinskii,Rubakov:1982df,Fabbri:1983us, 
Allen:1987bk, Sahni:1990tx}.
Detection of the CMB polarization pattern induced by
inflationary gravitational waves of wavelengths
comparable to the horizon has become a goal of next-generation
CMB experiments
\cite{Kamionkowski:1996ks, Kamionkowski:1996zd,
Zaldarriaga:1996xe, Seljak:1996gy,Cabella:2004mk}.  And now,
direct detection of the inflationary gravitational wave background (IGWB) with future spaced-based
gravitational-wave detectors at deci-Hertz frequencies has become
the subject of serious study \cite{BBO,Seto01}.

Detection of a gravitational-wave background, at either CMB or
direct-detection frequencies, would constitute a ``smoking gun'' for
inflation.  Moreover, since the amplitude of the IGWB
is determined by the energy scale of inflation at the time that
the relevant distance scale exited the horizon during inflation,
detection would provide important information about the
new ultra-high-energy physics responsible for inflation
\cite{Kamionkowski:1997av,Jaffe:2000yt}.  Since the
frequencies probed by the CMB and by direct detection are
separated by 16 orders of magnitude, the combination of both
provides a large lever arm with which the shape of the
inflaton potential can be constrained.

In this paper, we survey a range of inflationary models to
investigate the detectability of the IGWB with
satellite experiments, like NASA's Big Bang Observer (BBO) \cite{BBO} and
the Japanese Deci-Hertz Interferometer Gravitational-wave
Observatory (DECIGO) \cite{Seto01}, currently under study.  We restrict our
attention to slow-roll inflation models that are consistent with
measurements from the CMB and large-scale structure.  We show
how measurements of the IGWB amplitude at both
CMB and direct-detection scales can be used to constrain the
inflationary parameter space.

Previous work \cite{Barkana:94,Turner:1997,Ungarelli:05}
on direct detection of the IGWB has taken the
gravitational-wave spectrum to be a pure power law, considered
chaotic inflation \cite{Liddle:1994a,Battye:1996} or the IGWB due to a broken scale 
invariant potential \cite{Polarski:1999fb}.  In this
paper, we consider a wider range
of inflationary models (in the spirit of
Refs. \cite{Dodelson:1997hr,Kinney98}), and we solve the
inflationary dynamics
to go beyond the assumption of power-law power spectra.  With
this more accurate analysis, we find that for the forms of the 
inflaton potential considered here the direct detection of
the IGWB can break degeneracies between distinct inflationary
models that produce the same slow-roll parameters at
CMB/large-scale-structure scales for broken scale invariant potentials.

This paper is organized as follows.  In Section II, we review the
basics and relevant ingredients of inflation as well as
constraints from the CMB and large-scale structure (LSS) to the
inflationary observables.  We also discuss
the transfer function that relates the current
gravitational-wave amplitude to its primordial value. 
In Section III, we discuss the sensitivities of planned
space-based gravitational-wave observatories to the IGWB.  In
Section IV, we show for four different families of slow-roll
inflation models the IGWB amplitude and spectral
index $n_t$ at BBO/DECIGO scales that are allowed by current
CMB/LSS constraints to the $n_s$--$r$ parameter space (where
$n_s$ is the scalar spectral index and $r$ the tensor-to-scalar
ratio).  Section V compares the results of our calculations with
those obtained by extrapolating the power-law power spectra
inferred from CMB/LSS measurements to BBO/DECIGO scales.  
In section VI we discuss a family of slow-roll inflation models where the observational signature
in CMB/LSS studies for two different models is the same but differs
for the direct detection of the IGWB at BBO/DECIGO scales.  In
Section VII, we summarize and make some concluding remarks.

\section{Inflationary Dynamics and Perturbations
\label{inflation intoduction}}

\subsection{Homogeneous Evolution}

Inflation occurs when the cosmological expansion accelerates;
i.e., when $\ddot{a} > 0$, where $a(t)$ is the scale factor, and
the overdot denotes a derivative with respect to time $t$.  The
evolution of the scale factor is determined by the Friedmann
equation,
\begin{equation}
     H^2 \equiv \left( \frac{\dot a}{a} \right)^2 = \frac{8
     \pi}{3 M_{\mathrm{Pl}}^2} \rho -
     \frac{K}{a^2}, \label{eq:Friedmann}
\end{equation}
the continuity equation, $\dot{\rho} + 3H(\rho + P) = 0$,
and an equation of state $P(\rho)$, where $H$ is the Hubble
parameter, $\rho$ is the total energy density, $P$ is the pressure,
$\Mpl$ is the Planck mass, and $K$ is a constant related to the
3-space curvature.  From Eq.~(\ref{eq:Friedmann}) and the
continuity equation follows the ``acceleration'' equation,
\begin{equation}
     \frac{\ddot{a}}{a} = - \frac{4 \pi}{3 M_{\mathrm{Pl}}^2} (\rho+ 3 P).
\end{equation}
For an equation of state of the form $P = w \rho$,
where $w$ is a constant, inflation occurs when $w < -1/3$.

Consider now a spatially homogeneous scalar field $\phi$, the
``inflaton''.  It has an energy density and pressure,
\begin{eqnarray}
\rho &=& \frac{1}{2}\dot{\phi}^2 + V(\phi), \\
P &=& \frac{1}{2} \dot{\phi}^2 - V(\phi),
\end{eqnarray}
from which it follows that inflation occurs if $V(\phi) >
\dot{\phi}^2$.

The equation of motion for the inflaton is given by $\ddot{\phi}
+ 3 H \dot{\phi}+  V^{\prime} = 0$,
where the prime denotes differentiation with respect to $\phi$.
We assume that inflation has been proceeding for a long time
before any observable scales have exited the horizon, and so for
our purposes, the energy density is dominated by the inflaton
during inflation, the curvature term, $K/a^2$, is negligible as compared
to the inflaton energy density, and the evolution of the inflaton has been
attracted to the slow-roll regime (e.g., Ref.~\cite{Liddle94}).  If so, the evolution of the
inflaton and the scale factor are uniquely determined by $V(\phi)$.
Within the slow-roll approximation, the evolution is described
by the usual slow-roll parameters,
\begin{eqnarray}
     \epsilon &\equiv& \frac{M_{\mathrm{Pl}}^2}{16 \pi}
     \left(\frac{V^{\prime}}{V}\right)^2, \\
     \eta &\equiv&  \frac{M_{\mathrm{Pl}}^2}{8 \pi}
     \frac{V^{\prime \prime}}{V}, \\
     \xi &\equiv& \frac{M_{\mathrm{Pl}}^4}{64 \pi^2}
     \frac{V^{\prime} V^{\prime \prime \prime}}{V^2},
\end{eqnarray}
which are required to be small compared with unity for the
slow-roll approximation to be valid.  Toward the end of
inflation, $\epsilon$ grows, and inflation ends when
$\epsilon\simeq1$.  This statement can be made precise by the use of
``Hubble slow-roll'' parameters \cite{Liddle94}.

\subsection{Perturbations}

To leading order in the slow-roll approximation, the amplitudes
of the power spectra for density perturbations (scalar ``s'' metric
perturbations) and gravitational waves (tensor ``t'' metric
perturbations) can be written (e.g., Refs.~\cite{Stewart93,Lidsey97})
\begin{eqnarray}
     P_{s}(k) &\approx& \frac{128
     \pi}{3 M_{\mathrm{Pl}}^6}\frac{V^3}{V^{\prime 2}}
     \Bigg|_{k=aH}, \label{eq:scalar amp}\\
     P_{t}(k) &\approx& \frac{128}{3}\frac{V}{
     M_{\mathrm{Pl}}^4} \Bigg|_{k=aH},
     \label{eq:tensor amp}
\end{eqnarray}
as a function of wavenumber $k$, where $V$ and $V'$ are
evaluated when the relevant scale exits the horizon during
inflation.
The power spectra can be expanded in power laws,
\begin{eqnarray}
     P_s(k) &\approx& P_s(k_0)
     \left(\frac{k}{k_0}\right)^{1-n_s+(\alpha_s/2) \ln(k/k_0)}, 
\label{eq:Pts} \\
     P_t(k) &\approx& P_t(k_0)
     \left(\frac{k}{k_0}\right)^{n_t+(\alpha_t/2) \ln(k/k_0)},
\label{eq:Ptk}
\end{eqnarray}
where $k_0$ is a pivot wavenumber at which the spectral
parameters (e.g., Ref.~\cite{Liddle00}),
\begin{eqnarray}
     n_s (k) &\simeq& 1-6\epsilon+ 2 \eta, \label{eq:spectral} \\
     n_t(k) &\simeq& -2 \epsilon, \label{eq: tensor tilt}
     \\
     \alpha_s(k) &\simeq& 16 \epsilon \eta - 24 \epsilon^2 - 2 \xi,
     \\
     \alpha_t(k) &\simeq& 4 \epsilon \eta - 8 \epsilon^2,
\end{eqnarray}
are to be evaluated.  To a first approximation, the power
spectra are power laws with power-law indices $n_s$ and $n_t$,
although these indices may ``run''slightly with $k$, with a
running parameterized by $\alpha_s$ and $\alpha_t$ \cite{Kosowsky95}.
Finally, the tensor-to-scalar ratio is
\begin{equation}
     r \equiv \frac{P_t(k)}{P_s(k)} = 16
     \epsilon. \label{eq:consistency}
\end{equation}
In this paper, we will generally evaluate $P_s$, $n_s$, and
$\alpha_s$ at the distance scales of the CMB and large-scale
structure (LSS), where they are measured or
constrained.  In the Figures below, the tensor spectral index
$n_t$ will be evaluated at the distance scale relevant for
direct detection of gravitational waves.

\subsection{Number of $e$-Foldings \label{sec:e-foldings}}

The number of $e$-foldings of expansion between the
time, determined by $k= a_k H_k$, when a comoving distance scale
labeled by $k$ exited the horizon during inflation, and the end
of inflation is $N(k) \equiv
\ln\left(a_{\mathrm{end}}/a_k\right)$,
where $a_{\mathrm{end}}$ is the scale factor at the end of
inflation.  This must be (e.g., Ref.~\cite{Liddle:1993fq}),
\begin{eqnarray}
     &&N(k) = \label{num} \\
     &&62 - \ln \frac{k}{a_0 H_0} - \ln \frac{10^{16}\
     \mathrm{GeV}}{V_{k}^{1/4}}+ \ln
     \frac{V_k^{1/4}}{V_{\mathrm{end}}^{1/4}} - \frac{1}{3}
     \ln
     \frac{V_{\mathrm{end}}^{1/4}}{\rho_{\mathrm{rh}}^{1/4}},
     \nonumber
\end{eqnarray}
where $V_k$ is the inflaton potential when the comoving
scale $k$ crossed the inflationary horizon; $V_{\mathrm{end}}$
is the value of the potential at the end of inflation; and
$\rho_{\mathrm{rh}}  \sim T_{\mathrm{rh}}^4$ is the energy
density of the universe once radiation domination begins.  Here
we have assumed that the Universe was matter dominated after
inflation and before reheating and  that the
transition between radiation and matter domination is instantaneous.
In terms of the inflaton potential, the number
of $e$-foldings between two field values, $\phi_{i}$ and
$\phi_{f}$, is
\begin{equation}
     N(\phi_{i},\phi_{f}) \approx \frac{8
     \pi}{M_{\mathrm{Pl}}^2} \int_{\phi_f}^{\phi_i}
     \frac{V(\phi)}{V'(\phi)}d \phi,
\label{eqn:numberofefolds}
\end{equation}
where we have supposed the potential increases as the field increases so that the
field rolls towards the origin. 
Furthermore, if the potential determines a field value at which
inflation ends, we can
combine this equation with Eq.~(\ref{num}) and
$\rho_{\mathrm{rh}}$ to identify the field value when the
current Hubble volume exited the inflationary horizon.

If we use $10^{16}$ GeV for all the densities in
Eq.~(\ref{num}), we require 62 $e$-folds of inflation between
the time the current horizon distance exited the horizon during
inflation and the end of inflation.  The strength of the IGWB
is proportional to the inflaton-potential height
[cf., Eq.~(\ref{eq:tensor amp})], and as we will see,
detectability requires $V \gtrsim10^{15}$ GeV.  We will also see
that $V_{\mathrm{end}}$ is never much smaller than $V_k$.  Thus,
the only ratio in Eq.~(\ref{num}) that might be large is the
last.  Conservatively, the reheat temperature must be
$\gtrsim$ 1 MeV to preserve the successes of big-bang
nucleosynthesis, implying a lower limit of $N(a_0 H_0)\gtrsim
47$.  
This lower limit is significant, as the ratio of
gravitational-wave frequencies probed by the CMB/LSS (corresponding
to $k\simeq0.05\,\mathrm{Mpc}^{-1}$) and direct
detection is $\sim e^{35}$.  Therefore, if inflation results in
an IGWB in the ballpark of detectability by the CMB,
then inflation will last long enough to ensure the production of
gravitational waves on BBO/DECIGO scales (although it does not
necessarily guarantee a detectable amplitude).

\subsection{Constraints to Inflationary Observables \label{sec:constraints}}

We would like to survey only those inflationary models that are
consistent with current data.  Measurements of the
``inflationary observables''---i.e., the scalar and tensor
power-spectrum amplitudes, spectral indices, and running---come
from CMB measurements that probe  scales from the current Hubble
distance ($\sim 10^4$ Mpc) to $\sim 10$ Mpc scales, galaxy
surveys that constrain the matter power spectrum from
$600\ \mathrm{Mpc}$ down to $20\ \mathrm{Mpc}$, and from the
Lyman-$\alpha$ forest, which probes down to $\sim$ 1 Mpc
\cite{Croft:1998,Mandelbaum03}.  Constraints to the classical
cosmological parameters (e.g., the Hubble parameter, the
deceleration parameter, the baryon density, the matter density)
from other measurements
help limit the range of plausible values for the inflationary
observables that come from CMB/LSS measurements.

\begin{figure}[!h]
\centerline{\epsfig{file=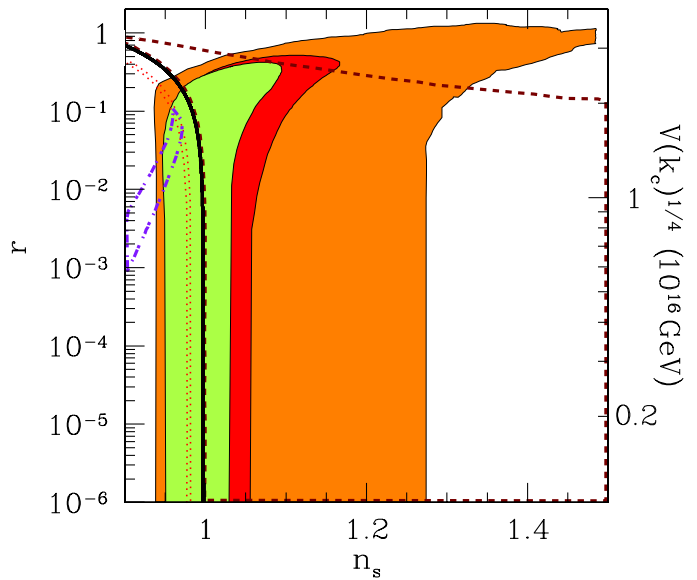,height=18pc,angle=0}}
\caption{Regions in the $n_s$--$r$ parameter space consistent
with the CMB-only (orange/medium gray) \cite{Peiris03}, CMB plus galaxy
surveys (red/dark gray), and CMB plus galaxy
surveys plus Lyman-alpha-forest constraints (green/light gray)
\cite{Seljak04}.  Here, $r$ is the
tensor-to-scalar ratio, and $n_s$ is the scalar spectral index
at CMB scales.  We plot on top of these regions the parameter
space occupied by the four
models of inflation we consider: 
power-law (solid black line), chaotic (dotted red), symmetry-breaking 
(dashed-dot purple) and hybrid (short-dashed dark-red).  The
right axis shows the energy scale $[V(k_c)]^{1/4}$ of inflaton.
}
\label{fig:nsr}
\end{figure}

The precise constraints to the inflationary observables
depend in detail on the combination of observational data sets.
In our discussions, we simply take as conservative ranges
$P_s(k_0) = (2.45 \pm 0.23) \times 10^{-9}$, $n_s = 1.0 \pm 
0.1$, and $|\alpha_s| < 0.04$ at a pivot wavenumber $k_0 = 0.05 \
\mathrm{Mpc}^{-1}$ \cite{Seljak04}.  We note that the
CMB-only constraints correspond to a pivot wavenumber, 
$k_0 = 0.002  \ \mathrm{Mpc}^{-1}$ \cite{Peiris03}. 
The errors we quote are not to be
interpreted as statistical errors bars, except in the case of
the value for $P_s(k_0)$, where the error quoted is $1\sigma$; 
rather, they simply
indicate a range of parameters that are in reasonable
concordance with measurements
\cite{Bennett03,Peiris03,Tegmark04b,Seljak04} and the range of
values that we use here.  In our discussion, we take
a conservative upper limit $r\lesssim1$ to the tensor-to-scalar
ratio.  The numerical results we show in Fig.~\ref{fig:all}
use slow-roll parameters consistent with the regions in the
$n_s$--$r$ parameter space, shown in Fig. \ref{fig:nsr}, taken
from analyses of CMB data alone, CMB plus galaxy surveys, and
CMB, galaxies, and the Lyman-alpha forest \cite{Peiris03,Seljak04}.

\subsection{Gravitational-Wave Transfer Function}

The gravitational-wave power spectrum $P_t(k)$ provides the
variance $\langle |h_k|^2 \rangle$ of the gravitational-wave
amplitude $h_k$ as that mode enters the horizon.  Once the
wavelength is smaller than the horizon, the gravitational wave
begins to oscillate, and its energy density $\rho_k \sim k^2
h_k^2$ redshifts $\propto a^{-4}$ like radiation.  It follows
that the gravitational-wave amplitude today is $h_k(t_0) =
h_k(t_k) (a_k/a_0)$, where $t_0$ is the time today, $t_k$
is the time of horizon entry, and $a_k=a(t_k)$ \footnote{This
can be shown more rigorously by solving the equation of motion
for $h$ during radiation domination and then comparing the
oscillation amplitude at late times with the initial amplitude.}.
During radiation domination (RD), $H \propto
a^{-2}$, so that $k \propto 1/a_k$, and during matter domination (MD), $H \propto
a^{-3/2}$, so that $k \propto 1/a_k^{1/2}$.  From these relations, we
find that the value of $h_k$ today scales with $k$ as
\begin{eqnarray}
     h_k &\propto& k^{-1} \ \mathrm{(horizon\ entry\ during\ RD)}, \\
     h_k &\propto& k^{-2} \ \mathrm{(horizon\ entry\ during\ MD)}.
\end{eqnarray}
Calculations (e.g., Refs.~\cite{Turner93,Pritchard:2004qp}) of the
transfer function intended for CMB predictions evolve the
wave equation more carefully through matter-radiation equality,
but the direct-detection frequencies are so high that the
scalings we have used here are fairly precise.  The sensitivities of
BBO/DECIGO will peak near a frequency 0.1 Hz, or wavenumber
$k=6.47\times 10^{13}\, \mathrm{Mpc}^{-1}$.  Matter-radiation
equality corresponds to $k_{\mathrm{eq}} \approx 0.05 h^2 \,
\mathrm{Mpc}^{-1}$.  
Therefore, the primordial gravitational waves observed by
the planned gravitational-wave observatories entered the horizon \emph{well}
before matter/radiation equality.  In fact, the modes that
entered the horizon during big-bang nucleosynthesis (at $T\sim 1$ MeV)
have frequencies $\sim 10^{-11}$ Hz.
Therefore, the gravitational waves probed by BBO/DECIGO are
$\sim 10$ orders of magnitude smaller than those associated with
big-bang nucleosynthesis (BBN) and must have
entered the horizon at temperatures $T\sim 10^7$ GeV.

These high temperatures imply a small correction to previous
calculations, which assumed $T\propto a^{-1}$, of the transfer
function due to the fact that it is actually $g_{*S}(T) a^3 T^3$
that remains constant, where $g_{*S}(T)$ is the effective number
of relativistic degrees of freedom contributing to the entropy
density.  Recalling that the Hubble parameter is $H\simeq 1.66\,
g_*^{1/2} T^2/M_{\mathrm{Pl}}$ during radiation domination
(where $g_*$ is the effective number of relativistic degrees of
freedom contributing to the energy density; at these
temperatures, $g_*\simeq g_{*S}$), the condition for horizon
entry for a physical wavenumber today 
($a_0 k= a_k H_k$), becomes
\begin{equation}
      1.66\, g_*^{1/2}(T_k) (T_0 T_k/M_{\mathrm{Pl}})
      \left(\frac{g_{* S}(T_0)}{g_{*
      S}(T_k)}\right)^{1/3}  = k.
\end{equation}
Today, $g_{*S}(T_0) = 3.91$ for photons and three species of massless 
neutrinos.
We then have,
\begin{equation}
      \left(\frac{T_k}{3.8 \times 10^6\, \mathrm{GeV}}\right) \left[
     \frac{g_{*}(T_k)}{100} \right]^{1/6} =
     \left(\frac{k}{6.5\times10^{13}\, \mathrm{Mpc}^{-1}}\right).
\end{equation}
Taking only standard-model particles, $g_{*}(T) \approx 100$ (or
roughly double if there is low-energy supersymmetry) and should
be roughly  independent of temperature.  We thus find that for
BBO/DECIGO scales,
\begin{equation}
     {\cal T}(k) = \frac{a_k}{a_0} = 2.1 \times 10^{-20}
     \left(\frac{k}{6.5\times 10^{13} \mathrm{Mpc}^{-1}}\right)^{-1}
     g_{100}^{-1/6}, \\
\end{equation}
where $g_{100} \equiv g_{*}(T_k)/100$.
With this transfer function, the root-variance of the IGWB today is
\begin{equation}
     \langle |h_k|^2 \rangle^{1/2} = P_t(k)^{1/2} \mathcal{T}(k).
\end{equation}
Free streaming of neutrinos, which occurs after neutrinos
decouple at a temperature $\sim 1 \ \mathrm{MeV}$, contributes
an anisotropic stress \cite{Weinberg04,Bashinsky:2005tv}.
However, the resulting
damping will be negligible for modes that enter the horizon so
much earlier than neutrino free-streaming.

If we define the logarithmic GW contribution to the critical density,
\begin{equation}
     \Omega_{\mathrm{GW}}(k) \equiv \frac{1}{\rho_c} \frac{d
     \rho_{\mathrm{GW}}}{d\ln k},
\end{equation}
then
\cite{Maggiore00},
\begin{equation}
     \Omega_{\mathrm{GW}}(k) h^2 = \frac{c^2 k^2 h^2}{6 H_0^2}
     \langle |h_k|^2 \rangle \equiv A_{\mathrm{GW}} P_t(k),
\end{equation}
where $H_0 \equiv 100\,h\ \mathrm{km}\,\mathrm{s}^{-1} \,
\mathrm{Mpc}^{-1}$, and $A_{\mathrm{GW}} = 2.74 \times 10^{-6}\,
g_{100}^{-1/3}$.  
In slow-roll inflation, $n_t<0$, so if we take
$P_s(k_0) \simeq 2.45\times 10^{-9}$ and $r\lesssim 1$
(corresponding to $V^{1/4} \lesssim 3.36\times 10^{16}$ GeV), we
find an upper limit $\Omega_{\mathrm{GW}} h^2 \lesssim 6.71 \times
10^{-15}g_{100}^{-1/3}$.

\section{Direct-Detection Thresholds \label{sec:transfer}}

Since the detection of a stochastic background of gravitational waves has to be
separated from the effect of sources of noise intrinsic to the detector, the
sensitivity to a stochastic background is different than the sensitivity to a
non-stochastic source.  The total stochastic signal in a given detector can be
written as a sum of a stochastic signal plus a stochastic noise,
$s(t) = h_n(t) + h(t)$.
Taking the Fourier transform of the total signal and considering the spectral
density of the noise and of the stochastic signal, we find that in order to
have a signal-to-noise greater than unity,
\begin{equation}
     \VEV{ |h(f)|^2}^{1/2} \gtrsim \left( \frac{2 f S_n(f)}{F}\right)^{1/2},
\end{equation}
where $F$ is a filling factor that accounts for the fact that a primordial
stochastic background will be isotropic on the sky, but the detector
will only be sensitive to a fraction of the sky, while $S_n(f)$ is the
spectral density associated with the detector noise.  

For omni-directional interferometers, such as the Laser
Interferometer Space Antenna (LISA), $F = 2/5$.  There is a
great improvement in
sensitivity when the signal from two independent detectors can
be combined through a correlation analysis between the two
detectors \cite{Christensen:1992wi, Flanagan:1993ix, Maggiore00, Cornish:2001qi}. 
Such a correlation increases
the sensitivity to a stochastic background such that,
\begin{eqnarray}
     && \VEV{|h(f)|^2}^{1/2}   \gtrsim \\ 
     &&1.12 \times 10^{-2}
     \left( \frac{2 f S_n(f)}{F}\right)^{1/2}
     \left(\frac{\mathrm{Hz} }{\Delta f}\right)^{1/4}
     \left(\frac{\mathrm{yr}}{\Delta T}\right)^{1/4},
     \label{eq:correlated}  
     \nonumber
\end{eqnarray}
where $\Delta f$ is the bandwidth over which the signals can be
correlated and $\Delta T$ is the integration time.
For a correlation analysis, the increase in sensitivity is under the assumption
that the detector noises are independent between the
two detectors, while the only correlation expected is due to stochastic signals
such as inflation.  For the single detector, the minimum
observable strain is independent of the integration time,
while for a correlation analysis, long-term observations are essential.
While LISA will not allow an opportunity for such a correlation
analysis, some mission concept studies for NASA's
Big Bang Observer (BBO) and Japan's Deci-Hertz Interferometer Gravitational
Wave Observatory (DECIGO) consider two (or more) systems such that the
improvement related to the correlation analysis can be exploited.  The design
for LISA currently places the sensitivity at approximately
$\Omega_{\mathrm{GW}} h^2 \sim 10^{-11}$.  Current designs for BBO place the
sensitivity of a single detector at $\Omega_{\mathrm{GW}} h^2
\sim 10^{-13}$ and
the sensitivity of a correlated extension at $\Omega_{\mathrm{GW}} h^2 \sim
10^{-17}$.  Finally, the ultimate goal for DECIGO is a sensitivity to
$\Omega_{\mathrm{GW}} h^2 \sim 10^{-20}$ \cite{Seto01},
corresponding to $V^{1/4} \sim 1.15\times 10^{15}$ GeV ($r \sim 10^{-6}$).

Besides a sensitivity to a stochastic background, one must also be concerned
about sources of a stochastic background, other than inflation.  One such
source is the background in extragalactic supernovae \cite{Buonanno04}.  Such
sources have the potential to wash out any signal that would otherwise be
observed from a primordial source, but the characterization of the amplitude
and frequency dependence of these sources is still uncertain.  Other sources 
of cosmological
gravitational-wave backgrounds
are white-dwarf/white-dwarf binaries \cite{Farmer:2003pa}, neutron-star/neutron-star 
binaries \cite{Schneider:2000sg} and neutron-star/white-dwarf binaries \cite{Cooray:2004ae}. 

\begin{figure*}
\begin{center}
\leavevmode
\epsfig{file=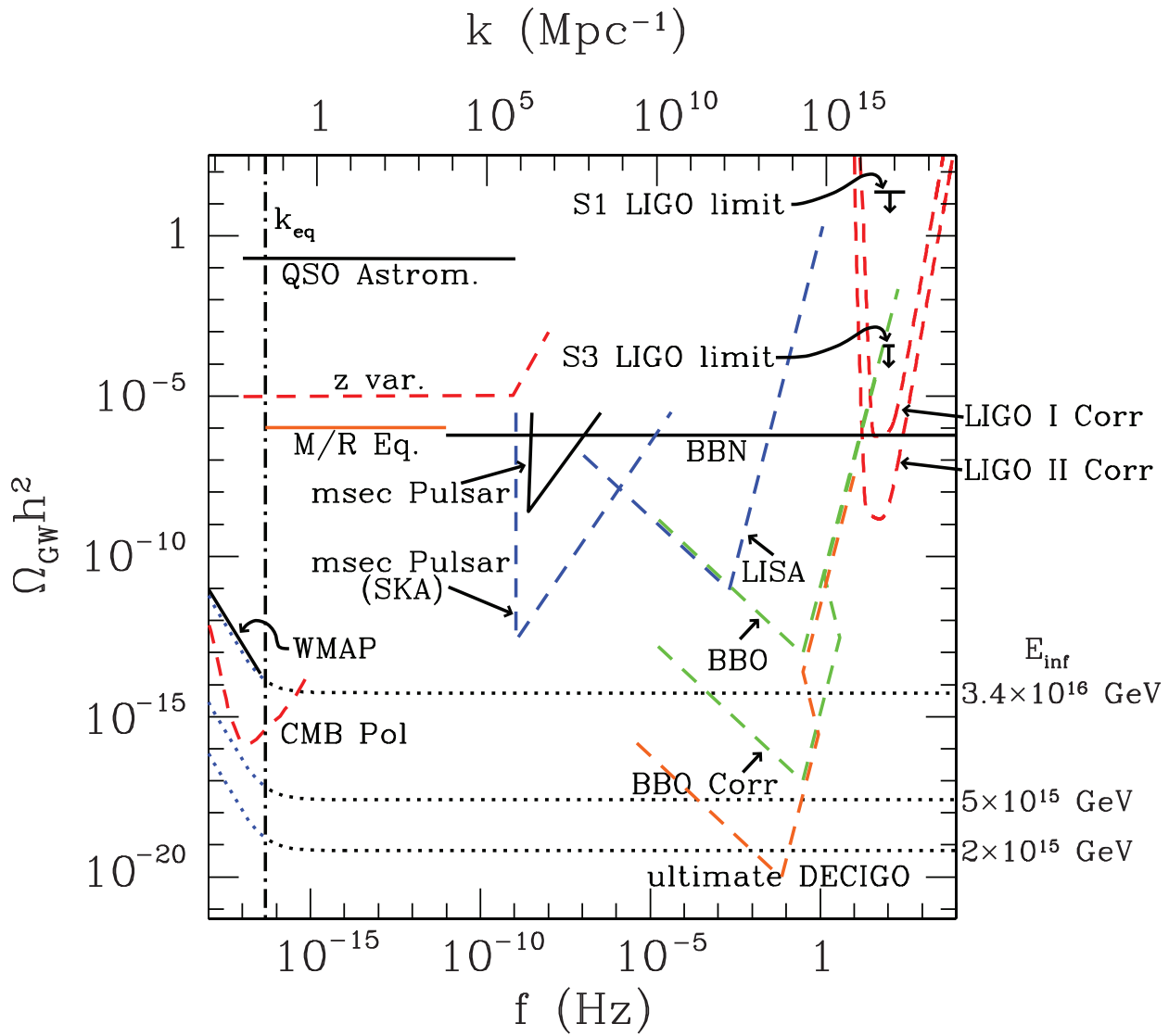,height=35pc,angle=0}
\end{center}
\caption{Current limits and projected sensitivities to a
stochastic gravitational-wave background versus the
gravitational-wave frequency.  The solid curves all indicate
current upper limits, while the dashed curves
indicate projected sensitivities. The LISA curve is
from Ref.~\cite{shanelarson} and BBO correlated from
Ref.~\cite{Buonanno04}.  The BBO sensitivity is estimated by
increasing the BBO-correlated curve by 4 orders of
magnitude [see Eq.~(\protect\ref{eq:correlated})].  The BBN
constraint  results from the limit to the number of relativistic
degrees of freedom at big-bang nucleosynthesis (e.g.,
Ref.~\cite{Allen:1997}); the
``M/R'' constraint is from CMB/LSS constraints to
matter-radiation equality \cite{Tristan}; the ``z. var''
curve is from Ref.~\cite{Seto:2005tq}; and the quasar-astrometry
limit from Refs.~\cite{Pyne:1995iy,Gwinn:1996gv}.  We note 
that the BBN and ``M/R'' constraints assume a scale invariant gravitational-wave
background that extends $\sim$ 60 $e$-folds below the current Hubble horizon.  LIGO 
sensitivities, taken from the LIGO Scientific Collaboration White Paper on Detector Research and Development \cite{LSC_whitepaper} are given in terms of a correlated analysis between the Hanford, WA and 
Livingston, LA sites [see Eq.~(\protect\ref{eq:correlated})].  The
run 1 LIGO limit (``S1 LIGO'') is from
Ref.~\cite{Abbott:2003hr} and the run 3 LIGO limit (``S3 LIGO'') is from Ref.~\cite{Abbott:2005ez}.
Also shown
are millisecond-pulsar timing constraints (current
\cite{Kaspi:1994hp,Lommen:2002je} and sensitivities projected
for the Square-Kilometer Array \cite{Kramer:2004rw}).
Curves corresponding to scale-invariant (i.e., $n_t=0$)
gravitational-wave backgrounds are shown (dotted curves), labeled by the
associated inflationary energy scales at CMB/LSS scales (but keep in mind that
slow-roll inflation generically predicts $n_t<0$, less power on
small scales).  The CMB/LSS currently constrains
this value to be below $3.36 \times 10^{16}$
GeV at CMB/LSS scales.  Future CMB measurements may be able to reach
energy scales near $10^{15}$ GeV
\protect\cite{Kesden:2002ku,Knox02,Seljak04b,Sigurdson:2005cp}.}
\label{fig:sensitivities}
\end{figure*}

Fig.~\ref{fig:sensitivities} shows the sensitivities to a
stochastic gravitational-wave background as a function of
frequency $f$ for a variety of gravitational-wave detectors.
Also shown are various current limits (as solid curves) as well
as a variety of projected direct and indirect sensitivities
(dashed
curves), and scale-invariant spectra parameterized by an
energy scale $V^{1/4}$ of inflation (dotted curves).  We also show limits from
current CMB experiments as well as the sensitivities expected
for future CMB-polarization experiments currently under study.

\section{BBO/DECIGO Amplitudes}

In this Section, we calculate the gravitational-wave amplitude
at BBO/DECIGO scales for several families of slow-roll
inflation models consistent with CMB/LSS constraints.

Measurements of the scalar amplitude $P_s$ and spectral index
$n_s$ at CMB/LSS scales, as well as upper limits to the tensor
contribution $r$ to
the CMB and to the running $\alpha_s$ of the spectral index, constrain the
inflaton potential and its derivatives at the field value
$\phi_c$ that corresponds to the time at which CMB/LSS scales
$k_c$ exited the horizon.  To be precise, we use $k_c=
0.05\,{\mathrm{Mpc}^{-1}}$.  In this work, we take as the
nominal BBO/DECIGO frequency $f=0.1$ Hz, corresponding to $k =
6.47\times10^{13} \, {\mathrm{Mpc}^{-1}}$ (and we note that
$\Omega_{\rm GW}(k)h^2\simeq$ constant for $n_t\simeq0$, so our
results will not depend too sensitively on the precise value of
$f$ we use).  CMB/LSS and BBO/DECIGO scales are therefore
separated by $\Delta N=\ln(6\times10^{13}/0.05)\simeq35$ $e$-folds of
inflation \footnote{We note that the actual expression that
relates two field values corresponding to known length-scales is
not given by Eq.~(\ref{eqn:numberofefolds}), which ignores, in part, the
variation of $H$ during inflation.  Instead, the exact
expression is given by, $$\ln{\left(\frac{k_1}{k_0}\right)} =
\sqrt{\frac{4\pi}{\Mpl^2}} \int_{\phi_1}^{\phi_0}
\frac{1-\epsilon}{\sqrt{\epsilon}} d\phi.$$  The error in our
expression is expected to be small, since we are only
considering the epoch of inflation far from its end,
so that we can always take $\epsilon \ll 1$, in 
which case the above expression becomes approximately equivalent
to Eq.~(\ref{eqn:numberofefolds}).}.
Eq.~(\ref{eqn:numberofefolds}) can then be used to
find the field value $\phi_g$ at the time that BBO/DECIGO
scales exited the horizon.

\subsection{Power-Law Inflation}

In power-law inflation, the inflaton potential takes the form,
\begin{equation}
     V(\phi) = V_0 e^{-p \phi/M_{\mathrm{Pl}}}.
\end{equation}
Power-law inflation is so called because the scale factor is a
power law $a(t) \propto t^{16\pi/p^2}$, and the Hubble parameter also
evolves as a power of time $t$.  The resulting scalar and tensor
power spectra are then pure power laws, with no running of the
spectral indices.  The parameter $\epsilon = p^2/(16\pi)$
always, so that inflation must be ended artificially at some
$\phi_{\mathrm{end}}$.
Although the potential has only two free parameters ($V_0$ and
$p$), there is an additional free parameter, namely, the value
of $\phi_{\mathrm{c}}$, which we are free to choose in this
particular family of models.  This model has also
$\eta=p^2/8\pi$, so $n_s=1-p^2/8\pi=1-2\epsilon$, and for
$n_s>0.9$ we find a constraint $\epsilon<0.05$.  The constraint
$r=16\epsilon \lesssim 1$ is comparable or a bit weaker.  Since
$n_s$ and $r$ depend in this model only on the parameter $p$,
these models occupy a curve in the $n_s$--$r$ parameters space,
which is indicated by the heavy solid curve
in Fig. \ref{fig:nsr}.  The constraint $\Delta N=35$ to the
number of $e$-folds between CMB/LSS and BBO/DECIGO scales tells
us that
\begin{equation}
     \Delta N= \frac{8\pi}{p}\frac{\phi_g-\phi_c}{M_{\mathrm{Pl}}}
     \simeq 35,
\end{equation}
from which it follows that
\begin{equation}
     \frac{P_t(k_g)}{P_t(k_c)} = \frac{V(\phi_g)}{V(\phi_c)} =
     e^{-(p^2/8\pi)\Delta N} = e^{-2\epsilon \Delta N}.
\end{equation}
We thus find that the gravitational-wave amplitude at DECIGO/BBO
scales is
\begin{eqnarray}
\Omega_{\mathrm{GW}} h^2 &=& A_{\rm GW} P_t(k_g) =
A_{\rm GW} r P_s(k_c) e^{-2\epsilon \Delta N} \nonumber \\
&=& 1.08\times 10^{-13} \epsilon \, e^{-70\, \epsilon
(\Delta N/35)}  \nonumber \\
&& \times\left( \frac{P_s(k_c)}{2.45\times10^{-9}}
\right) \left(\frac{A_{\rm GW}}{2.74\times10^{-6}} \right).
 \end{eqnarray}    
This expression is maximized for $\epsilon=1/(2\Delta N)\simeq 1/70$ at a
value $\Omega_{\rm GW}^{\rm max} h^2 = 5.68\times 10^{-16}$.
Interestingly enough, the IGWB detectability through direct
detection is maximized for relatively small $\epsilon$, while
the detectability with the CMB is maximized at larger
$\epsilon$ \cite{Liddle:1994a}.  Given that CMB sensitivities are expected to get to
$r\sim 0.01$ in the relatively near future with the CMB, and
then to $r\sim 10^{-4}$ with a next-generation satellite
experiment, it is unlikely that this model would produce a
direct detection without producing a detectable CMB signal.  

\begin{figure*}
\begin{center}       
\leavevmode       
\epsfig{file=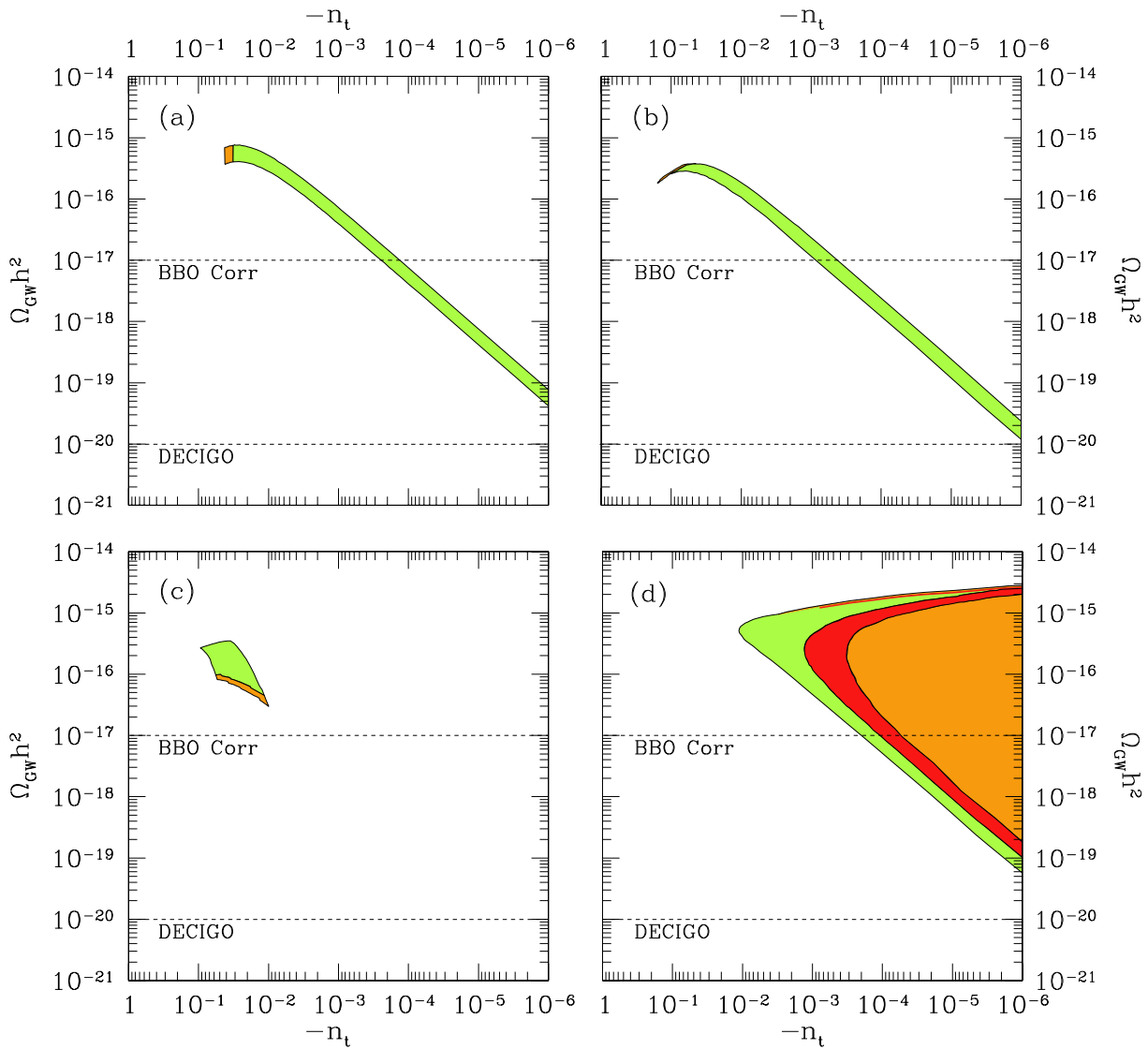,height=35pc,angle=0}
\end{center}     
\caption{Regions in the $\Omega_{\rm GW} h^2$--$n_t$ parameter
space for (a) power-law, (b) chaotic,
(c) symmetry-breaking, and (d) hybrid inflation.  The colored (shaded)
regions map out the corresponding regions in
Fig. \protect\ref{fig:nsr}.  Here, the
gravitational-wave density $\Omega_{\rm GW} h^2$ and spectral
index $n_t$ are both evaluated at DECIGO/BBO scales.  Also shown
are the sensitivity goals of BBO and DECIGO.}
\label{fig:all}
\end{figure*} 

Fig.~\ref{fig:all}(a) shows the region of the $\Omega_{\mathrm{GW}}
h^2$--$n_t$ parameter space (at BBO/DECIGO scales) that the
$n_s$--$r$ parameter space shown in Fig.~\ref{fig:nsr} maps to
for power-law inflation.  The breadth in $\Omega_{\mathrm{GW}}
h^2$ of the region is due to the 30\% (at $3\sigma$) uncertainty
in $P_s(k_c)$.  If power-law inflation is the correct model of
inflation, then the IGWB is directly detectable with BBO for
$r \gtrsim 10^{-3}$ and $r \gtrsim10^{-6}$ with DECIGO.

\subsection{Chaotic Inflation \label{sec:chaotic}}

In chaotic inflation, the inflaton potential is,
\begin{equation}
     V(\phi) = V_0 \left(
     \frac{\phi}{M_{\mathrm{Pl}}}\right)^{\alpha}.
\end{equation}
In this family of models, $\epsilon(\phi) = (\alpha^2 / 16\,
\pi) (\Mpl/\phi)^2$, and therefore inflation ends when
$\phi=\phi_{\mathrm{end}} \equiv \alpha \Mpl/(4 
\sqrt{\pi})$.  If there are $N_c$ $e$-folds of inflation between
CMB horizon exit and the end of inflation, then
Eq.~(\ref{eqn:numberofefolds}) gives us $\phi_c^2 =
(\Mpl^2/16\pi)(4\alpha N_c + \alpha^2)$.  We also have
$\eta(\phi)=\alpha(\alpha-1)(\Mpl/\phi)^2/(8\pi)$ from which
it follows that at CMB/LSS scales,
\begin{equation}
     n_s = 1 - 2 \frac{\alpha +2 }{4 N_c + \alpha}.
\end{equation}
Noting that $47\lesssim N_c \lesssim 62$, the constraint $n_s
>0.9$ gives us a constraint $\alpha \lesssim 4 N_c/ 19 - 40/19$.  
The constraint on the tensor-to-scalar ratio,
$r=16\epsilon = 16\, \alpha / (4 N_c +\alpha) \lesssim 1$,
leads to a slightly less stringent limit, $\alpha \lesssim 4 N_c/15$.  
We note that the scalar running $\alpha_s=
- 2(1-n_s)^2/(2+\alpha)$ is always within the current
observational constraints since $|1-n_s| \lesssim 0.1$.  
This family of models is thus
parameterized by two parameters: $47\lesssim N_c \lesssim 62$
and $\alpha \lesssim 4 N_c/ 19 - 40/19$.  Note that each choice of
$(\alpha,N_c)$ maps onto a point in the $n_s$--$r$ parameter
space, so we could just as well choose $n_s$ and $r$ as our two
independent parameters.  If we choose to do so, then we assign
$N_c$ and $\alpha$ by $N_c = (1-\epsilon)/(1-n_s-2\epsilon)$ and
$\alpha = 4 \epsilon/(1-n_s-2\epsilon)$, where $\epsilon=r/16$.

For a fixed value of $N_c$, this family of models is represented
by a curve in the $n_s$--$r$ parameter space; a spread in the
range of values for $N_c$ broadens this curve into a region in
the $n_s$--$r$ parameter space, as indicated by the region
enclosed by the dotted red curves in Fig. \ref{fig:nsr}

Once $\alpha$ and $N_c$ are specified, the potential prefactor
is fixed by
\begin{equation}
     V_0 =  \frac{3 \alpha^2 P_s(k_c)}{128\, \pi} \left(
     \frac{16\, \pi}{4\, \alpha N_c+ \alpha^2}
     \right)^{(\alpha+2)/2}\Mpl^4.
\end{equation}
The gravitational-wave amplitude at direct-detection scales is
then simply,
\begin{equation}
     \Omega_{\rm GW} h^2 = \frac{128}{3} A_{\rm GW}
     \frac{V_0}{\Mpl^4} \left( \frac{\phi_g}{\Mpl}
     \right)^\alpha,
\end{equation}
where the field value $\phi_g$ at the time direct-detection
scales undergo horizon crossing is given by
\begin{equation}
     \phi_g^2 = \frac{\Mpl^2}{16\pi} (4\alpha N_g +\alpha^2),
\end{equation}
and where $N_g=N_c-35 \equiv N_c-\Delta N$ is the number of
$e$-folds before the end of inflation that DECIGO/BBO scales
exit the horizon.
For chaotic inflation, the value of $n_t$ at
DECIGO/BBO scales will differ from (and generally be larger in
amplitude, or more negative than) that at CMB/LSS scales.  The 
value of $n_t$ at DEICGO/BBO scales will differ from that at CMB/LSS
scales; it will be given by $n_t(\phi_g) = -2 \epsilon(\phi_g)$.  

Fig.~\ref{fig:all}(b) shows the region of the $\Omega_{\mathrm{GW}}
h^2$--$n_t$ parameter space (at BBO/DECIGO scales) that the
$n_s$--$r$ parameter space shown in Fig.~\ref{fig:nsr} maps to
for chaotic inflation.  The breadth in $\Omega_{\mathrm{GW}}
h^2$ of the region is due to the spread in the $\alpha$--$N_c$
parameter space for {\it fixed} $P_s(k_c)=2.45\times10^{-9}$;
there will be a slight additional vertical broadening beyond
that shown due to the uncertainty in this parameter.
If chaotic inflation is the correct model of 
inflation, then the IGWB is directly detectable with BBO for
$r \gtrsim 10^{-3}$ and $r \gtrsim10^{-6}$ with DECIGO.

\subsection{Symmetry Breaking Inflation}

We now consider the Higgs potential,
\begin{equation}
     V(\phi) = V_0 \left[1-\left(\frac{\phi}{\nu}\right)^2\right]^2,
\end{equation}
parameterized by $V_0$ and a Higgs vacuum expectation value
$\nu$.  Our treatment of this family of models will parallel
that for chaotic inflation.  In these models, $\phi$ starts near
the origin and then rolls toward $\phi=\nu$.  The slow-roll
parameters are $\epsilon(\phi) = (\Mpl^2 \phi^2/4\pi \nu^4)
[1-(\phi/\nu)^2]^{-2}$, and $\eta(\phi) = (\Mpl^2/2\pi\nu^2) [3
(\phi/\nu)^2 -1] [1-(\phi/\nu)^2]^{-2}$, from which we infer an
end to inflation,
\begin{equation}
     \phi_{\mathrm{end}} = \left[\frac{\Mpl^2}{2\pi}\left(1+2 \pi
     \frac{\nu^2}{\Mpl^2} -  \sqrt{1+ 4 \pi
     \frac{\nu^2}{\Mpl^2}}\right)\right]^{1/2} .
\label{end symmetry}
\end{equation}
The field value at which CMB/LSS scales undergo horizon crossing
during inflation is
\begin{equation}
     \phi_c^2 = \frac{N_c \Mpl^2}{\pi} + \phi_{\mathrm{end}}^2 -
     2 \nu^2 \ln (\phi_{\mathrm{end}}/\phi_c).
\end{equation}
At CMB/LSS scales,
\begin{equation}
     n_s = 1-
     \frac{1}{\pi}\left(\frac{M_{\mathrm{Pl}}}{\nu}\right)^2
     \frac{1+ 3y_c^2}{\left(1-y_c^2\right)^2}, 
\label{eq:nS symmetry}
\end{equation}
where $y_c\equiv \phi_c/\nu$.  Since $n_s$ is a decreasing
function of $y$, the constraint $n_s>0.9$ requires $\nu\gtrsim1.8\,
\Mpl$.  The prefactor $V_0$ is then fixed by the constraint,
\begin{equation}
     V_0 = \frac{3}{ 8 \pi} P_s(k_c) (\Mpl/\nu)^2 \frac{ y_c^2
     }{(1-y_c^2)^4} \Mpl^4.
\end{equation}
Once this normalization is
fixed, these models are parameterized by $\nu$ and $N_c$, and
$n_s$ and $r$ are fixed once these two parameters are
specified.  As in chaotic inflation, we may alternatively take
as our two free parameters $n_s$ and $r$, and then determine
$\nu$ and $N_c$, although the inversion is not as tractable
algebraically as in chaotic inflation.

The gravitational-wave amplitude at direct-detection scales is
then simply,
\begin{equation}
     \Omega_{\rm GW} h^2 = \frac{128}{3} A_{\rm GW}
     \frac{V_0}{\Mpl^4} \left( 1-y_g^2 \right)^2,
\end{equation}
where $y_g \equiv \phi_g/\nu$, and the field value $\phi_g$ at
the time direct-detection scales undergo horizon crossing is
given by
\begin{equation}
     \phi_g^2 = \frac{N_g \Mpl^2}{\pi} + \phi_{\mathrm{end}}^2 - 2 \nu^2 \ln (\phi_{\mathrm
     {end}}/\phi_g),
\end{equation}
and where again, $N_g=N_c-35 \equiv N_c-\Delta N$ is the number of
$e$-folds before the end of inflation and the time when DECIGO/BBO scales
exit the horizon.  The value of $n_t$ at DECIGO/BBO scales will
differ from that at CMB/LSS scales; it will be given by $n_t(\phi_g)
= -2\epsilon(\phi_g)$.
We also note that the running of the scalar spectral index at
CMB/LSS scales is
\begin{equation}
     \alpha_s = -\frac{1}{\pi^2}\left(\frac{M_{\mathrm{Pl}}}{\nu}\right)^4
     y_c^2 \frac{5+3y_c^2}{(1-y_c^2)^4}.
\end{equation}
We check in our numerical results that all of the models we
consider are consistent with the bound to this parameter, $|\alpha_s| < 0.04$.
In particular we find that $|\alpha_s| \lesssim 10^{-3}$. 

Fig.~\ref{fig:all}(c) shows the region of the $\Omega_{\mathrm{GW}}
h^2$--$n_t$ parameter space (at BBO/DECIGO scales) that the
$n_s$--$r$ parameter space shown in Fig.~\ref{fig:nsr} maps to
for symmetry-breaking inflation.  The breadth in $\Omega_{\mathrm{GW}}
h^2$ of the region is due to the spread in the $\nu$--$N_c$
parameter space for {\it fixed} $P_s(k_c)=2.45\times10^{-9}$.
If symmetry-breaking inflation is the correct model of 
inflation, then the IGWB will be detectable with BBO and
DECIGO.  Incidentally, we have also investigated potentials of
the form $V(\phi)=V_0[1-(\phi/\nu)^p]^2$ with $p > 2$ \cite{Kinney:1995cc}.  In these
models, the symmetry-breaking scale can be reduced below $\Mpl$,
however the IGWB amplitude is then reduced below the level
accessible to BBO/DECIGO for $\nu \lesssim 0.1 \Mpl$.

\subsection{Hybrid Inflation \label{hybrid}}

Hybrid inflation generally requires two scalar fields
\cite{Linde94}, but the phenomenology can be modeled by a single
scalar field with the potential,
\begin{equation}
V(\phi) = V_0\left[1+\left(\frac{\phi}{\mu}\right)^2 \right],
\end{equation}
and the selection of a value for $\phi_{\mathrm{end}}$ (with the only requirement that 
$\phi_{\mathrm{end}} > 0$).  We note that this form for the potential is not to be taken to 
be generic within the class of hybrid inflation but only as a particular example.  Other 
forms exist, such as those found in, e.g., Refs.~\cite{Dvali:1994ms, Linde:1997sj, Lyth:1996kt}. 
Defining $y \equiv \phi/\mu$, we find that the slow-roll
parameters are given by,
$\epsilon(y) = (\Mpl/4 \pi \mu^2) y^2[1+y^2]^{-2}$, and 
$\eta(y) = (\Mpl/4 \pi \mu^2)[1+y^2]^{-1}$.  In these models,
$\epsilon$ is maximized at $y=1$ with a value less than unity if
$\mu > \Mpl/(4\sqrt{\pi})$.  For smaller
values of $\mu$, inflation ends soon after $y=1$
\cite{Copeland:1994}\footnote{One can numerically determine that
within a fraction of an $e$-folding $y \rightarrow 0$ and
therefore must pass through $y_{\mathrm{end}}$.}.  The dynamics
of these models resembles those of chaotic inflation, which we
have already considered, and so we consider them no further.
New inflationary dynamics arises for $\mu>\Mpl/(4\sqrt{\pi})$ and $y \leq 1$, and so
we restrict our attention here to this regime.

The field value at which CMB/LSS scales undergo
horizon crossing during inflation is, 
\begin{equation}
     y_c^2 = 2 \ln \left(\frac{y_c}{y_{\mathrm{end}}}\right) -
     y_{\mathrm{end}}^2 - \frac{N_c}{2\pi}
     \left(\frac{\Mpl}{\mu}\right)^2. 
\label{eq:ychybrid}
\end{equation}
Since $y_c$ is taken to be a free parameter,
Eq.~(\ref{eq:ychybrid}), along with 
$47 \lesssim N_c \lesssim 62$, determines the value of $y_{\mathrm{end}}$. 
From the slow-roll parameters,
\begin{equation}
     n_s = 1 + \frac{\Mpl^2}{2 \pi \mu^2} \frac{1-2
     y_c^2}{(1+y_c^2)^2},
\label{eq:nShybrid}
\end{equation}
at CMB/LSS scales.
The above expression for $n_s$ is maximized at $y=0$, and at
this field value becomes $n_s = 1 + \Mpl^2/2\pi\mu^2$, which
shows that we can have $n_s > 1$ in hybrid inflation.
The pre-factor $V_0$ is then fixed by the constraint, 
\begin{equation}
V_0 = \frac{3}{32 \pi} \left(\frac{M_{\mathrm{Pl}}}{\mu}\right)^2
P_s(k_{\mathrm{c}}) \frac{y_{\mathrm{c}}^2}{(1+y_{\mathrm{c}}^2)^3} \Mpl^4.
\end{equation}
Once this normalization is fixed, these models are parameterized by $\mu$ and
$y_c$, and $n_s$ and $r$ are fixed once these two parameters are specified.  As in chaotic 
inflation, we may alternatively take as our two free parameters $n_s$ and $r$, and 
then determine $\mu$
and $y_c$.  In particular, these can be determined from 
\begin{eqnarray}
y_c &=& \left[\frac{r}{8(n_s-1)+2r}\right]^{1/2}, \\
\mu &=& 2 \left[ \frac{2}{\pi} \frac{4(n_s
-1)+r}{\left[8(n_s-1)+3r\right]^2}\right]^{1/2} \Mpl.
\end{eqnarray}

The gravitational-wave amplitude at direct-detection scales is
then simply given by,
\begin{equation}
\Omega_{\mathrm{GW}} h^2 = 
\frac{128}{3} A_{\mathrm{GW}} \frac{V_0}{\Mpl^4}(1+y_g^2),
\end{equation}
where the field value $y_g$ at the time direct-detection scales
undergo horizon crossing is given by
\begin{equation}
y_g^2 = 2 \ln \left(\frac{y_g}{y_{\mathrm{end}}}\right) - 
y_{\mathrm{end}}^2 - \frac{N_g}{2\pi} \left(\frac{\Mpl}{\mu}\right)^2,
\end{equation}
and where again, $N_g = N_c -35 \equiv N_c-\Delta N$ is the number
of $e$-folds before the end of inflation and the time when
DECIGO/BBO scales exit the horizon.  The value of $n_t$ at
DECIGO/BBO scales will differ from that at CMB/LSS scales; it
will be given by $n_t(y_g) = -2 \epsilon(y_g)$.  
The running of the tensor spectral index,
\begin{equation}
     \alpha_t(y) = \frac{1}{4 \pi^2} \left(\frac{M_{\mathrm{Pl}}}{\mu}\right)^4
\frac{y^2(1-y^2)}{(1+y^2)^4},
\label{running hybrid}
\end{equation}
can be positive in this class of models.
Thus, for $y<1$, $\alpha_t > 0$, and
the running is positive, indicating that as $y$ evolves, the
tensor spectral index becomes less negative.  As we have seen in
the previous models, a non-negligible gravitational-wave
amplitude at CMB/LSS scales leads to a ``large'' amplitude at
direct-detection scales primarily due to a small,
negative, tensor spectral index.  We therefore expect this model to
produce the largest gravitational-wave amplitude at
direct-detection scales.
We also note that the running of the scalar spectral index at
CMB/LSS scales is
\begin{equation}
     \alpha_s = \frac{\Mpl^4}{2 \pi^2 \mu^4}
     \frac{y_c^2(y_c^2-2)}{(1+y_c^2)^4}.
\end{equation}
With the restriction that $y_c \leq 1$, $\alpha_s$ is maximized
at $y = \sqrt{2-\sqrt{3}}$.  We note from this
that if $\mu \geq 0.69 \Mpl$ the observational bound on $\alpha_s$ 
is satisfied for all $y_c \leq 1$.  For $\mu$ not satisfying
this restriction, there will be some range of $y_c$ which are
incompatible with observations.  This restriction is taken into
account in our numerical calculations. 

Fig.~\ref{fig:all}(d) shows the region of the $\Omega_{\mathrm{GW}}
h^2$--$n_t$ parameter space (at BBO/DECIGO scales) that the
$n_s$--$r$ parameter space shown in Fig.~\ref{fig:nsr} maps to
for symmetry-breaking inflation.  The breadth in $\Omega_{\mathrm{GW}}
h^2$ of the region is due to the spread in the $\mu$--$N_c$
parameter space for {\it fixed} $P_s(k_c)=2.45\times10^{-9}$.
If hybrid inflation is the correct model of 
inflation, then the IGWB may be detectable with BBO and
DECIGO, but it is not guaranteed. 

\section{The (running) power-law approximation}

During inflation, the value $\phi$ of the scalar field can change
very little as the scale factor $a(t)$ grows extremely rapidly.
It is therefore a feature of inflation that a vast range of
distance scales can correspond to a small change in $\phi$.
This motivates the power-law
expansions (with a slight running of the spectral index) in
Eqs.~(\ref{eq:Pts}) and (\ref{eq:Ptk}), which assume that the
inflaton potential can be accurately approximated by its Taylor
expansion (to second order) about a given inflaton value.  These 
power-law expansions are particularly appropriate when studying the
CMB and large-scale structure (e.g.,
Refs. \cite{Peiris03,Seljak04,Ungarelli:05}), which involve a spread in
distance scales of maybe three orders of magnitude.

\begin{figure*}
\begin{center}       
\leavevmode       
\epsfig{file=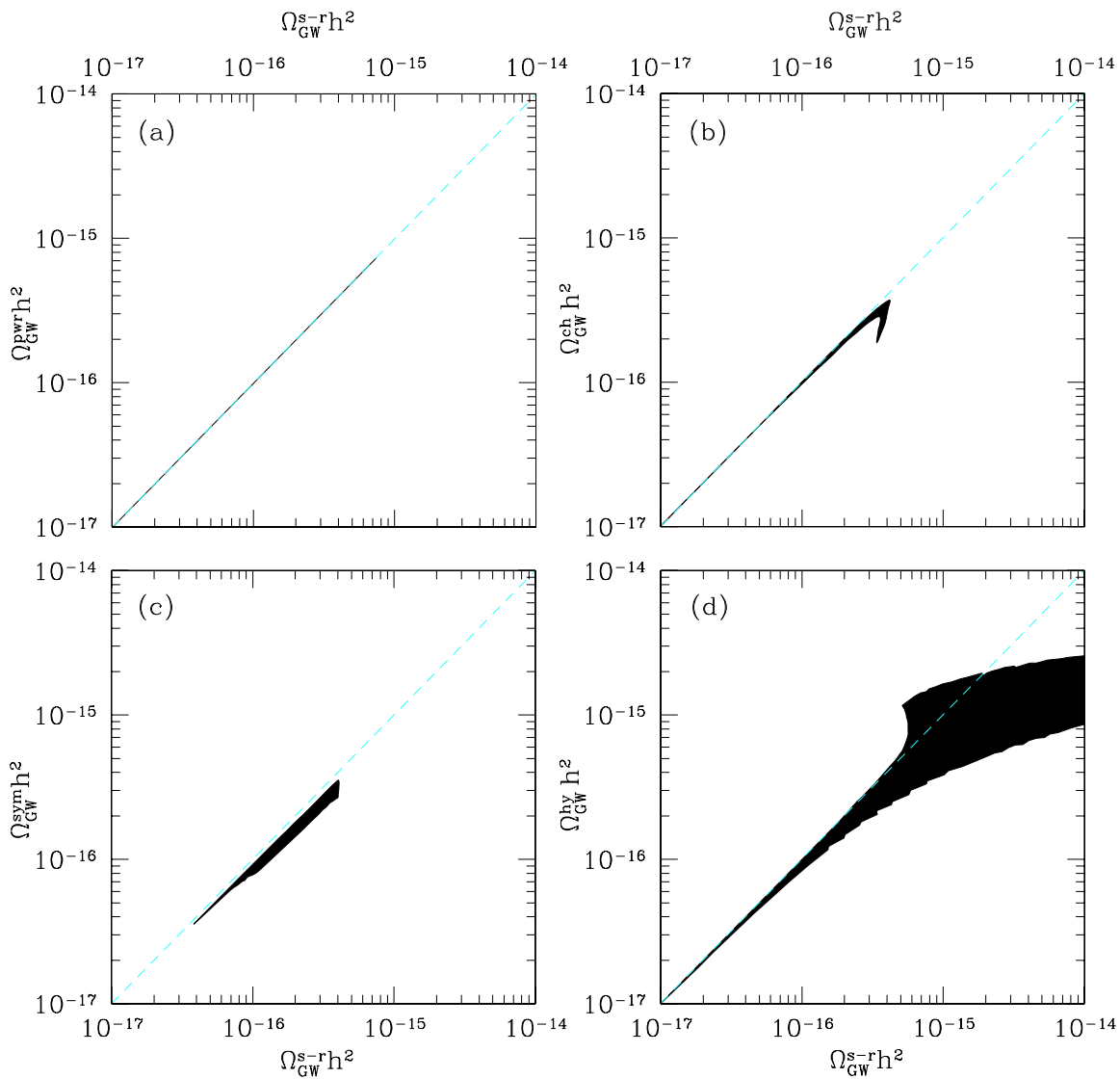,height=35pc,angle=0}
\end{center}     
\caption{Here we plot the inflationary gravitational-wave
background amplitude $\Omega_{\mathrm{GW}}h^2$ obtained with the
exact inflationary dynamics described in Section IV versus the
amplitudes obtained with the power-law approximations (with
slow-roll parameters fixed by CMB/LSS observations) given in
Section IIb.  The panels show results for (a) power-law, (b)
chaotic, (c) symmetry-breaking, and (d) hybrid inflation.  The
regions are models taken from the $n_s$--$r$ parameter space in
Fig. \ref{fig:nsr} (and shown in Fig. \ref{fig:all}) and the
blue dashed curves indicate equality.}
\label{fig:slowroll}
\end{figure*} 

However, BBO/DECIGO frequencies are separated from those probed
by the CMB/LSS by roughly sixteen orders of magnitude.  The
inflaton may thus traverse a significant distance, and so it
is not obvious that the Taylor expansion approximation that underlies
the power-law approximation (even with the running of the
spectral index) will remain valid.  For example, in Eqs.~(\ref{eq:spectral}) and (\ref{eq: tensor tilt}), 
the tensor and scalar tilt are written in
terms of the first-order slow-roll parameters, while second- and higher-order
corrections (e.g., Ref. \cite{Stewart:1993bc}) may be
important when extending the power spectrum over large physical
scales. Similarly, one must also account for higher-order derivatives of
the tilt, beyond the running considered with $\alpha_s$ and $\alpha_t$.
For the calculation performed here, higher order corrections are not
important as $\Omega_{\mathrm{GW}}h^2$  was directly determined with model parameters
describing the inflaton potential, rather than through the power spectrum.
Assuming the Taylor approximation
is valid, then measurements of $P_s$, $n_s$,
and $r$ at CMB/LSS scales fix the parameters $P_t(k_c)$, $n_t$,
and $\alpha_t$ in Eq. (\ref{eq:Ptk}), which can then be used
to predict $P_t(k_g) \propto \Omega_{\mathrm{GW}} h^2$, the IGWB
amplitude at BBO/DECIGO scales. 
An approach based on the Taylor expansion was considered in  Ref. \cite{Ungarelli:05} 
to estimate the GW amplitude at frequencies corresponding to direct
detections. 
Fig.~\ref{fig:slowroll} plots the exact IGWB amplitude
obtained from the calculation in the previous Section versus
that obtained from the power-law approximation.  For small IGWB
amplitudes, $r=16\epsilon \rightarrow 0$, and so the potentials
are very close to flat and the power law tends to be a good
approximation, and is indeed a good approximation for the four
classes of models we have considered.  For power-law inflation,
where the power spectra are precisely power laws, the two
results are identical.  For chaotic and symmetry-breaking
inflation, $\epsilon$ becomes large when the IGWB amplitude
becomes large, and $\epsilon$ evolves during inflation in such a
way that the power-law approximation overestimates the true IGWB
amplitude at direct-detection scales.  The behavior of hybrid
inflation is a bit more subtle.  The running of the tensor
spectral index is $\alpha_t = (r/8) (n_s-1+r/8)$ which, for
$n_s> 1-r/8$ can be positive.  The (running) power-law would
then suggest that $n_t$ will become positive at some small
distance scale, which cannot be [see Eq.~(\ref{eq: tensor tilt})].  
The power-law approximation
can then overestimate the true IGWB amplitude at BBO/DECIGO
scales.  On the other hand, in hybrid inflation, unlike chaotic
and symmetry-breaking inflation, $\epsilon$ can indeed decrease
as inflation proceeds, and so the direct-detection amplitude may
also be underestimated by the power-law approximation.  Both
behaviors are seen in Fig. \ref{fig:slowroll}.

It should also be kept in mind that the tensor spectral index
$n_t$ is most generally different at CMB scales than it is at
direct-detection scales, and it is conceivably measurable at
both.  Determination of $n_t$ at both distance scales could
therefore distinguish between inflationary models.  For example,
in power-law inflation, $n_t$ remains precisely constant, while it
can change by roughly a factor of two for chaotic inflation for
models with a directly detectable IGWB.  Realistically, though,
the tensor spectral indices are generically (although not in full
generality) small, and so running of the tensor spectral index
will be difficult to measure.

Finally, the four classes of models we have considered are not
at all exhaustive, and another inflaton potential could yield a
direct-detection IGWB amplitude different from those we
have considered here and different from what extrapolation from
CMB/LSS would suggest from the power-law approximation.  For
example, in models with broken scale invariance
\cite{Starobinsky:1992ts,Lesgourgues:1997en}, the direct-detection amplitude could
be considerably different.

\section{Broken Scale Invariant Spectrum}

To demonstrate that direct observations 
of the IGWB can distinguish between different forms of 
the inflaton potential, we consider as a toy model
the broken scale invariant (BSI) potential, which features a sharp change 
in the slope of the inflaton potential 
at some transition scale \cite{Starobinsky:1992ts,Lesgourgues:1997en,Polarski:1999fb}.  Such models have been invoked to explain, e.g., the paucity of dwarf galaxies
 observed around the Milky Way \cite{Kamionkowski:1999vp, Zentner:2002xt, Zentner:2003yd}.  

Consider a potential of the form, 
\begin{equation}
V(\phi) = V_0 \times \begin{cases} (1+A \phi), \mathrm{\phi \gtrsim 0} \\
	(1+c A \phi),\mathrm{\phi \lesssim 0}, \end{cases} \label{eq:linear_pot}
\end{equation}
where $V_0$ is the overall normalization, $A$ is the slope 
of the potential at CMB/LSS scales and $c$ parameterizes how the slope
changes after the break at $\phi = 0$. 
We allow $\phi_c$ to be a free parameter, only requiring that it
be before the break in the potential at $\phi = 0$.  This freedom supposes that field value
at which inflation ends is not necessarily determined by the form of the potential
in Eq.~(\ref{eq:linear_pot}).  
In order to choose $\phi_c$, we place 
the break (i.e., $\phi = 0$) $N_0$ $e$-folds from $\phi_c$.  A natural
choice for $N_0$ is 10, since
 we suppose that CMB/LSS scales constrain the inflaton 
potential from $10^4\  \mathrm{Mpc}$ to $1\ \mathrm{Mpc}$.  
The normalization of the scalar 
power-spectrum then fixes the normalization $V_0$
of the potential through the expression, 
\begin{equation}
V_0 = \frac{3 P_s(k_c)}{128} \frac{A^2 \Mpl^2}{(1+A \phi_c)^3}\Mpl^4.
\end{equation}
We then integrate Eq.~(\ref{eqn:numberofefolds}), 
assuming the transition at $\phi = 0$ has a negligible contribution,
between $\phi_c$ and $\phi_g$ with $N = 35$ 
in order to find $\phi_g$.  We require that inflation not end before we reach $\phi_g$.  
For $V(\phi)$ as in Eq.~(\ref{eq:linear_pot}), we find
that inflation ends soon after the field reaches 
a value, $\phi_* = \Mpl/(4\sqrt{\pi})-(c A)^{-1} $.
This then places a constraint on the combination $c A$, 
\begin{equation}
c A < \frac{4 \sqrt{\pi}}{\Mpl \sqrt{1+4(N_g-N_0)}}. \label{eq:Ac1}
\end{equation}

At CMB/LSS scales we find that $\eta(\phi_c) = 0$, and
\begin{equation}
\epsilon(\phi_c) = \frac{A^2 \Mpl^2}{4 A^2 \Mpl^2 N_0 + 16\pi}.
\end{equation}
From the above expression and Eqs.~(\ref{eq:spectral}) and (\ref{eq:consistency}), we can see explicitly that $n_s$ and $r$ only 
depend on our choice of $N_0$ and $A$. 
At BBO/DECIGO scales we find
\begin{equation}
\Omega_{\mathrm{GW}} h^2 =  4 A_{\mathrm{GW}} P_s(k_c) A^2 \Mpl^2 \frac{\sqrt{A^2 c^2 \Mpl^2 
(N_0 - N_g)+4\pi}}{(A^2 \Mpl^2 N_0 + 4\pi)^{3/2}}, \label{eq:linear_omega}
\end{equation}
where $N_g$ is the number of $e$-foldings between $\phi_c$ and $\phi_g$.
In order for there to be $N_g$
$e$-folds between $\phi_c$ and $\phi_g$ the slope of the potential cannot be too large, requiring 
\begin{equation}
c A \leq \frac{2 \sqrt{\pi}}{\Mpl \sqrt{N_g - N_0}}. \label{eq:Ac2}
\end{equation}
Comparing this to Eq.~(\ref{eq:Ac1}), we find that this constraint is \emph{slightly} less restrictive.  
We can see
that this amplitude depends not only on $N_0$ and $A$ but also on $c$.  Therefore, 
potentials that share approximately the same Taylor expansion at CMB/LSS scales, but different 
expansions at BBO/DECIGO scales, will produce overlapping observations in the $(n_s,r)$ plane at CMB/LSS scales and different gravitational-wave amplitudes at BBO/DECIGO scales.  
With the constraint in Eq.~(\ref{eq:Ac1}), we find that as $c$ increases towards 
its maximum value (for a fixed $A$), the amplitude of the IGWB changes by an order of magnitude.  
For example, for $n_s = 0.9$ and $r = 0.27$ we have $1.0 \times 10^{-16} \lesssim 
\Omega_{\mathrm{GW}} h^2 \lesssim 1.0 \times 10^{-15}$; 
for $n_s = 0.99$ and $r = 3.16 \times 10^{-3}$ we have $2.1 \times 10^{-18} 
\lesssim \Omega_{\mathrm{GW}} h^2 \lesssim 2.1 \times 10^{-17}$; 
and for $n_s = 1.0$ and $r = 3.18 \times 10^{-5}$ we have $2.1 \times 10^{-20} \lesssim 
\Omega_{\mathrm{GW}} h^2 \lesssim 2.1 \times 10^{-19}$.

\section{Discussion}

In this paper we have calculated the gravitational-wave
amplitude at direct-detection scales for four classes of
inflationary potentials with parameters consistent with current
constraints from the CMB and LSS.  The gravitational-wave
amplitude $\Omega_{\mathrm{GW}}h^2$ is proportional to the
height $V(\phi_g)$ of the inflaton potential at the time that
direct-detection comoving scales exit the horizon.  Our current
theoretical understanding does not fix $V(\phi_g)$; it is
constrained to be $V^{1/4} \lesssim 3.4\times10^{16}$ GeV from the
CMB, and it could conceivably be as low as $T\sim 1$ MeV without
violating observational constraints.  Moreover, detectability
of the IGWB with BBO or DECIGO requires $V^{1/4} \gtrsim
10^{15}$ GeV, close to the upper allowed limit.  It thus seems,
{\it a priori}, that detectable models occupy a small region of
parameter space.

That said, however, there are indeed constraints to inflationary
models that come from the CMB and large-scale structure, notably
constraints to the density-perturbation amplitude and spectral
index.  Fig.~\ref{fig:all} indicates that when we go through the
exercise of writing down simple functional forms for the
inflationary potentials and imposing current constraints, there
are large regions of parameter space that lead to directly
detectable IGWB amplitudes.  In particular, for the
symmetry-breaking potential, which looks perhaps like the type
of Higgs potentials we might associated with grand unification,
current constraints lead to a directly detectable IGWB
amplitude.  

The promise of detectability traces back to the fact that
$\Omega_{\mathrm{ GW}} h^2 \propto V \propto (V')^{4/3}$, the
last proportionality tracing back to Eqs.~(\ref{eq:scalar amp})
and (\ref{eq:tensor amp}) for fixed density-perturbation
amplitude $P_s$.  Thus, if the potential is extremely flat, $V'
\rightarrow 0$, then the IGWB will be tiny.  However, if the
potential takes a form in which $V' \sim V/\phi$, which seems
theoretically natural, then the required density-perturbation
amplitude is achieved with $V\sim 10^{(15-16)}$ GeV,
the range that produces an accessible IGWB amplitude.

There is of course still plenty of room for inflation to be
correct and for the IGWB amplitude to be well below the BBO or
DECIGO threshold.  For example, in power-law inflation and
chaotic inflation, the IGWB amplitude becomes small when
$n_s\rightarrow 1$; i.e., when scale invariance is achieved
which, in these models corresponds to small $V'$.  On the other
hand, a value $n_s \rightarrow 1$ does not, more generally,
imply a small IGWB amplitude.  For example, in hybrid inflation
one can have $n_s=1$ if $y_c^2=1/2$ [cf.,
Eq. (\ref{eq:nShybrid})], and for $\mu\gtrsim 1.8\, \Mpl$, the
potential can reach values at CMB/LSS scales of $V\sim3\times
10^{16}$ GeV, which even after the decrease to BBO/DECIGO scales
remains within reach of detection, as shown in
Fig.~\ref{fig:all}(d).

There may of course be alternatives to inflation, such as cyclic
models \cite{Steinhardt02}
or the pre big-bang model
\cite{Veneziano:1991ek,Gasperini:1994xg,Vernizzi:2000vc,Veneziano00, Brustein:1995ah,
Buonanno:1996xc},
that have
completely different IGWB spectra.  Although the
cyclic model predicts a blue tensor spectrum, which might
improve detectability at small scales, BBN constrains the
amplitude of the gravitational-wave amplitude to be orders of
magnitude below the BBO/DECIGO sensitivities \cite{Boyle04}.

Our conclusion is that direct detection of the IGWB is unlikely
without detection with the CMB polarization.  Still, direct
detection could be extraordinarily valuable even if the IGWB is
detected first in the CMB.  Direct detection would provide yet
another cross-check that the curl component in the CMB
polarization is due to gravitational waves, as opposed to some
other mechanism (e.g., vector modes, cosmic shear, or
foregrounds).  Since it occurs on such vastly different distance
scales, direct detection can verify that it is a nearly
scale-invariant spectrum of gravitational waves, as predicted by
inflation, as opposed to some other phenomenon that might only
produce large-wavelength gravitational waves.  It would provide
evidence for the continuation of inflation to distance scales
well beyond those implied by the smoothness of the Universe
suggested by the successes of BBN.   The
large lever arm provides an opportunity to discriminate between
inflationary models that produce the same CMB/LSS observables.
Even within the
context of a given inflationary potential, the large lever arm
associated with direct detection may allow a measurement of
inflationary parameters that may be more precise than those
accessible with the CMB/LSS alone.  For example, an uncertainty
of $10\%$ in $\epsilon$ from the CMB/LSS translates to a $\sim
(10^{15})^{0.2}\sim 1000$ uncertainty in the BBO/DECIGO
amplitude.  Thus, a detection {\it alone}, with no better than
an order-unity amplitude uncertainty, corresponds to a
measurement of $\epsilon$ to roughly 0.02, probably better than
is accessible with the CMB/LSS alone.  Finally, the deci-Hertz
IGWB amplitude counts the number of relativistic
degrees of freedom at temperatures a bit above the electroweak
symmetry-breaking scale, and may thus be used to probe for new
degrees of freedom associated with supersymmetry of some other
new physics at the electroweak scale \cite{Seto:2003kc}.  The direct search for
inflationary gravitational waves may thus be warranted.

\acknowledgments
During the preparation of this paper, we learned of unpublished recent work, along similar lines, by Will Kinney as part of a BBO mission concept study \cite{BBO2}.  This work was supported in part by DoE
DE-FG03-92-ER40701 and NASA NNG05GF69G.  TLS acknowledges the
support of a NSF Graduate Fellowship.

\bibliography{bibliography}

\end{document}